\documentclass{article}
\begin{document}
\title{Twistors versus harmonics}
\author{C. Lovelace\\ Physics Dept., Rutgers University,\\
        Piscataway, NJ 08854-8019. \\ [9pt]
        lovelace @ physics.rutgers.edu }
\date{June 21, 2010}
\maketitle
\begin{abstract}
By lifting full Yang-Mills theory to $R_4 \times S_2\,$,
Mason et al. obtained MHV vertices by gauge
transformation. Their Lagrangian depended on long
intricate twistor manipulations. Spinor $S_2$ harmonics
give a one page proof, eliminating all the $CP_3$
apparatus.
\end{abstract}

\section{Introduction}                      % SECTION 1

It is fashionable to abandon Lagrangians completely and 
construct twistor\linebreak S matrices. However survivors
from the 1960's have unpleasant memories of a similar
fashion. Still interesting therefore are a series of
five papers by Mason et al. \cite{mason,boels7,boels8}.
They claimed that full Yang-Mills [YM] theory can be
lifted to $R_4 \times S_2$, and then GAUGE transformed
to an MHV basis. This ought to be much simpler than
Mansfield's canonical transformation \cite{mansfield}.
Unfortunately they wrote in a hieratic dialect (twish?)
dear to Oxford mathematicians. It is important to
find a similar lifted Lagrangian for supergravity
where previous attempts have foundered \cite{nair},
so a demotic translation would be useful. I will only
consider bosonic YM in Euclidean spacetime. Fermions
would just complicate the notation.

In two dimensions free wave equations can be solved
just by going to complex (or lightcone) coordinates.
In 4D this is ambiguous -- there are $S_2$ possibilities
and there is no reason to make the same choice at each
point. Thus one is led to consider $R_4 \times S_2$
in order to make analytic or lightcone gauge covariant
\cite{atiyah}. Twistor people regard $R_4 \times S_2
= C_2 \times CP_1$ as a stepping stone to
$CP_3$, but I will not pursue that. Instead of twistors
I will use the $S_2$ harmonics of Galperin et al.
\cite{harmonics}, which are a spinor extension of
familiar spherical harmonics. They were used long ago
\cite{selfdual} to solve selfdual YM. I will first
review this (\S 2,3), and then extend it to full YM
theory (\S 4). In \S 5 I check that the result is
equivalent to Mason's. After explaining prerequisites,
my proof fits into one page.

\section{Spinor harmonics}                          %SECTION 2

Here $x$ is a point in Euclidean spacetime; $u, v, w$
are points on the fiber $S_2$. Since Lagrangians are
local on the $R_4$ base, no fiber bundle theory is
needed. $D$ denotes an $su(2)$ generator on the fiber.
Otherwise partial derivatives are $\partial$. When gauge
potentials $A$ are added, I call both types $\nabla$.
The $\pm$ superscripts count a conserved charge on
the $S_2$ fiber.

In standard spinor notation \cite{penrose}, irreducible
representations of $Spin(4) = SU(2)_L \times SU(2)_R$
are $\psi_{(AB\cdots )(A'B'\cdots )}$, where ( ) means
symmetrize. Indices are raised or lowered by
$\psi^A = \epsilon^{AB} \psi_B,\; \psi_A = \psi^B 
\epsilon_{BA}$, where $\epsilon_{AB} = \epsilon^{AB} =
{0 \quad\; 1 \choose -1 \;\; 0} $.

Consider $su(2)_R$. We could represent it on $S_2$ by
the usual angular momentum operators $L_\pm ,
L_0$ expressed in polar coordinates $\theta ,\phi $.
However as shown by Galperin et al.\cite{harmonics},
it is more convenient to label a point $u \in S_2$
by spinors $u^{+A'},\; u^-_{B'}\;$, satisfying
(in the above sign convention)
\begin{equation}\label{a1}                     % a1
    u^{+A'}\;u^-_{A'}\;=\;1\;=\;-\,\epsilon^{B'A'}
                         \;u^+_{B'}\;u^-_{A'} \;,
\end{equation}
\begin{equation}\label{a2}                     % a2
    (u^{+A'})^* \;=\;-\, u^-_{A'}\;,
\end{equation}
where * means complex conjugate. Eqs.(\ref{a1})-(\ref{a2})
could be solved by a single complex number $u^+_1$,
as in stereographic projection, but they are more
powerful unsolved (to avoid patching). 
The $su(2)_R$ generators are then
\begin{equation}\label{a3}                     % a3
    D^{\pm\pm} \equiv L_\pm  \;=\;  u{^{\pm A'}}
                    \frac{\partial}{\partial u^{\mp A'}} ,
    \quad    D^0 \equiv \,2 L_0   \;=\;
         u^{+A'} \frac{\partial}{\partial u^{+A'}}
      -  u^{-A'} \frac{\partial}{\partial u^{-A'}} . 
\end{equation}
Since $S_2 = SO(3)/SO(2)$, the $D^0$ eigenvalue is
required to be a conserved charge. Its value is the
difference between the number of + and - superscripts.
(This reduces $su(2)$ representations to the coset space.)
If $u, v$ are two points on $S_2$, define
\begin{equation}\label{a4}                    % a4
    <uv>\;\;\equiv\;\;\epsilon^{A'B'}
                      \;u^+_{A'}\;v^+_{B'}\; .
\end{equation}
$<uv>$ has charge ++, but I will only show this
when it matters.
An important identity \cite{harmonics} is
\begin{equation}\label{a5}                    % a5
    <u|\,(D^{++})^{-1}\,|v>\;\;=\;\;<u v>^{-1}.
\end{equation}
This is the bilocal Green's function on $S_2$ that
inverts the differential operator $D^{++} \equiv L_+$
(subject of course to global conditions).

Euclidean spacetime coordinates $x^\mu$ can be converted
to bispinors by
\begin{equation}\label{a6}                       % a6
    x^{BB'} = \;i x^4 + (\vec{\sigma} .\,\vec{x} ),
\end{equation}
and then shuffled into analytic (+) and anti-analytic 
(-) doublets
\begin{equation}\label{a7}                       % a7
    y^{\pm B} \;\equiv\;  u^\pm_{B'} \;x^{BB'} .
\end{equation}
Similarly
\begin{equation}\label{a8}                       % a8
    \partial^\pm_B\;=\;u^{\pm B'}\;\partial_{BB'}\;.
\end{equation}
Note that by (\ref{a1}) the signs reverse
\begin{equation}\label{a9}                       % a9
    \partial^\pm_B \;=\;\pm\,
                   \partial /\partial y^{\mp B} .
\end{equation}
Now add gauge potentials (antihermitian matrices)
\begin{equation}\label{a10}                      % a10
    \nabla_{BB'}\;\equiv\;\partial_{BB'} + A_{BB'}\;.
\end{equation}
It is well known \cite{penrose} that
\begin{equation}\label{a11}                      % a11
    [\nabla_{BB'} \;,\; \nabla_{CC'}]  \;=\;
    F_{(BC)}\;\epsilon_{B'C'} 
                \;+\; F_{(B'C')}\;\epsilon_{BC}\;.
\end{equation}
This separates the field strength into selfdual
$F_{(BC)}$ and antiselfdual [ASD] \linebreak 
components. Thus far $x,\;u$ are independent. 
Multiplying by harmonics
\begin{equation}\label{a12}                      % a12
    \nabla^\pm_B\;\equiv\; u^{\pm B'}\;\nabla_{BB'}\;,
\end{equation}
(\ref{a11}) gives
\begin{equation}\label{a13}                      % a13
    [\nabla^+_B \;,\; \nabla^+_C]  \;=\;
    \epsilon_{BC}\;\, u^{+B'} u^{+C'}\;F_{(B'C')}\;,
\end{equation}
an equation which will be very useful later.

\section{Selfdual Yang-Mills}                  %  SECTION 3

Now suppose the YM theory is selfdual $F_{(B'C')} = 0$.
Here I am summarizing \cite{selfdual}. Then
\begin{equation}\label{a14}                        % a14
    [\nabla^+_B \;,\; \nabla^+_C] \;=\; 0 \;.
\end{equation}
So by Frobenius' theorem \cite{ward} there exists a
gauge transformation $\Lambda (x, u)$ that locally
flattens this connection. Since it depends on $u$, it
will unflatten the $su(2)_R$ algebra (\ref{a3}).
Thus OLD to NEW gauge transforms
\begin{equation}\label{a15}                        % a15
    \nabla^+_B \;\to\; \partial^+_B , \qquad
    D^{\pm\pm} \;\to\; \nabla^{\pm\pm}
    \;\equiv\; D^{\pm\pm} + A^{\pm\pm} .
\end{equation}
The two new components $A^{\pm\pm}$ are enough to
compensate for the two old components $A^+_B$,
so $D^0$ can stay flat.

By (\ref{a3}) and (\ref{a12}) in old gauge
\begin{equation}\label{a16}                        % a16
    [D^{++}\;,\; \nabla^+_B] \;=\; 0\;,
\end{equation}
so in new gauge
\begin{equation}\label{a17}                        % a17
    [\partial^+_B\;,\; \nabla^{++}] \;=\; 0 \;,
\end{equation}
which means by (\ref{a9}) that $A^{++}$ is independent
of $\;y^-_B\;$. The $su(2)_R$ algebra becomes in new gauge
\begin{equation}\label{a18}                        % a18
    [\nabla^{++}\;,\; \nabla^{--}] \;=\; D^0 \;,
\end{equation}
with no potential on the rhs. As explained in
\cite{selfdual}, this can be used to solve for $A^{--}$
in terms of $A^{++}$. The result is most easily described
by perturbing the Green's function (\ref{a5})
\pagebreak
\[                                                 % a19
    \mathcal{G}^{--} (u,v) \;\equiv\;\;
                      <u|\,[D^{++} + A^{++}]^{-1}\,|v>
\]
\begin{equation}\label{a19}
     =\; <uv>^{-1} \;-\; \int dw <uw>^{-1} A^{++}(x,w)
                     <wv>^{-1} \;+\; \ldots\ldots .
\end{equation}
The $n^{th}$ term of the perturbation expansion
has $n-1$ integrals over $S_2$, while $x$ 
stays fixed. Then by \cite{selfdual} eq.(III.26),
\begin{equation}\label{a20}                        % a20
    -\,A^{--} (x,u) \;=\; \lim_{v \to u}\,
           [\,\mathcal{G}^{--} (u,v) - <uv>^{-1}\;]\;.
\end{equation}
Because of the singular denominators this \emph{does} 
depend on $y^-_B \:$ .

Lastly we can find the nonzero part of the field strength
$F_{(BC)}$. In old gauge, (\ref{a3}) and (\ref{a12}) gave
\begin{equation}\label{a21}                        % a21
    \nabla^-_B \;=\; [D^{--}\;,\; \nabla^+_B\,]\;.
\end{equation}
In new gauge this becomes by (\ref{a15})
\begin{equation}\label{a22}                        % a22
    A^-_B  \;=\; - \partial^+_B \, A^{--} \;.
\end{equation}
In old gauge, (\ref{a11}) with $F_{(B'C')} = 0$ gave
by (\ref{a1}) and (\ref{a12}) 
\begin{equation}\label{a23}                        % a23
    [\nabla^+_B \;,\; \nabla^-_C ] \;=\;- F_{(BC)}\;.
\end{equation}
Thus in new gauge
\begin{equation}\label{a24}                        % a24
    F_{(BC)}\;=\;\partial^+_C\,\partial^+_B\,A^{--}\;.
\end{equation}
So the selfdual theory is entirely determined by the
unconstrained prepotential $A^{++}(y^+_B, u) $,
which is independent of $y^-_B$ by (\ref{a17}).
Of course similar results were obtained earlier by
Ward \cite{ward} using twistors, but they are not
as easy to follow as \cite{selfdual}.

\section{Full Yang-Mills}                       % SECTION 4

Let us see how rapidly this technique reproduces long
esoteric twistor manipulations. For completeness I will 
first put the usual YM Lagrangian into Chalmers-Siegel form.

Let $F^\pm$ be the SD/ASD YM field strengths as in
(\ref{a11}). It is well known \cite{atiyah,eguchi}
that $(F^+)^2 - (F^-)^2$ is a total divergence.
Therefore the standard YM Lagrangian (trace over
gauge group assumed)
\begin{equation}\label{b1}                        % b1 25
    \mathcal{L}\;=\;\frac{1}{16} F^{\mu\nu} F_{\mu\nu}
    \;=\; \frac{1}{8} \,[(F^+)^2 + (F^-)^2]
\end{equation}
is equivalent to
\begin{equation}\label{b2}                        % b2 26
    \mathcal{L} \;=\; \frac{1}{4}\, (F^-)^2
       \;\equiv\;\frac{1}{4}\,F^{(C'D')}\,F_{(C'D')}\;.
\end{equation}
It is convenient to introduce a dummy field
$G^{(C'D')} (x)$ , which can be functionally integrated
out by completing the square, leaving (\ref{b2}):
\begin{equation}\label{b3}                        % b3 27
    \mathcal{L}\;=\;\frac{1}{2}\,G^{(C'D')}\,F_{(C'D')}
            \;-\; \frac{1}{4}\,G^{(C'D')}\,G_{(C'D')}\;.
\end{equation}
If we omitted the second term, G would be a Lagrange
multiplier for selfdual YM \cite{chalmers}.

Now the race starts. By (\ref{a13})
\begin{equation}\label{b4}                        % b4 28
    [\nabla^+_C\;,\;\nabla^+_D ] \;=\; \epsilon_{CD}
              \;\; u^{+C'} u^{+D'} \,F_{(C'D')} \;.
\end{equation}
This suggests a natural way to include $G^{(C'D')}$.
Introduce a scalar potential \\[1pt]
$B^{--} (x,u)$ and identify
$G^{(C'D')}$ with three of its Fourier modes on $S_2$
\begin{equation}\label{b5}                        % b5 29
    G^{(C'D')}(x)\;=\;\int du
                 \;u^{+C'}u^{+D'}\,B^{--}(x,u)\;.
\end{equation}
Then the first term of (\ref{b3}) becomes
\begin{equation}\label{b6}                        % b6 30
    \frac{1}{2} \,G^{(C'D')}\, F_{(C'D')}  \;=\;
    \frac{1}{4} \int du\;B^{--}(x,u)\;\epsilon^{CD}
                \;[\,\nabla^+_C\;,\; \nabla^+_D\,]\;.
\end{equation}
At first sight the second term of (\ref{b3}) becomes
\begin{equation}\label{b7}                        % b7 31
    - \frac{1}{4} G^{(C'D')}\, G_{(C'D')}  \;=\;
    - \frac{1}{4} \int du \int dv \;B^{--} (x,u)
                  <\!u^+ v^+\!>^2 B^{--} (x,v)  \;,
\end{equation}
where $<\!u^+ v^+\!>$ is (\ref{a4}). 
However this is not gauge invariant. The gauge 
potentials $B(u),\; B(v)$ need to be connected 
by Wilson lines on $S_2$
\[
    \exp(\;\int_u^v dw \; A^{++}(w)\;)\;,
\]
forming a loop when traced. (In 2D we can fix
$A^{--}\;=\;0$.) This is equivalent to inserting
propagators (\ref{a19}), but then we need two more
$<\!u^+ v^+\!>$ factors to conserve the $D^0$
charge. So the unique correct formula is
\begin{equation}\label{b10}                      % b10 32
    -\frac{1}{4} \int du\int dv <\!u^+ v^+\!>^4\,
    Tr\{ B^{--} (x,u) \,\mathcal{G}^{--} (u,v)
       \,B^{--} (x,v) \,\mathcal{G}^{--} (v,u) \} \;,
\end{equation}
where
\begin{equation}\label{b11}                      % b11 33
    \mathcal{G}^{--}(u,v) \;\equiv\;\;
                <u|\,[D^{++} + A^{++}]^{-1}\,|v> .
\end{equation}
Check: if $A^{++}=0\,$, (\ref{b10}) = (\ref{b7}) 
by (\ref{a5}).
So far $A^{++}$ belongs to a trivial 2D gauge theory
independent of $x$.
We can also add a Lagrange multiplier term to implement
(\ref{a16}) :
\begin{equation}\label{b9}                        % b9 34
    B^{---C}\; [\,\nabla^{++}\;,\; \nabla^+_C \,]\; .
\end{equation}

Now perform a gauge transformation $\Lambda (x,u)$.
This will change 
\[
    A^{++}(u)\; \to \; A^{++}(x,u) \;,
\] 
but of course it won't eliminate
$A^+_C$ completely, unlike the SD case (\ref{a15}).
We can however set one component $A^+_2 = 0$,
thus defining a new gauge.
The final Lagrangian is 
\newpage
\[                                               % b12 35
    \frac{1}{4} \,\epsilon^{CD} \int du \;Tr\{
         B^{--}\; [\nabla^+_C\;,\;\nabla^+_D]\;\}
       + \int du \;Tr\{ B^{---C}\; 
             [\nabla^{++}\;,\; \nabla^+_C]\;\}
\]
\begin{equation}\label{b12}
    \\ - \frac{1}{4} \int du \int dv <uv>^4
    Tr\{ B^{--}(x,u)\,\mathcal{G}^{--}(u,v)\,
         B^{--}(x,v)\,\mathcal{G}^{--}(v,u) \}.
\end{equation}
In the gauge $\partial^-_C A^{++} = 0$ 
it reduces to (\ref{b3}) + (\ref{b5}).
In the gauge $A^+_2 = 0$ the unwanted $BAA$ vertex 
in the first term vanishes, and (\ref{a19}) gives
a series of MHV vertices instead. The bilocal 
$S_2$ term first appeared in \cite{old} .

\section{Comparison}                           % SECTION 5

The clearest of the five Mason et al. papers is
Ref.\cite{boels7}, where \cite{mason} are usefully
summarized. (My notations were chosen for
easy comparison). Boels starts with the usual 
twistor equation
\begin{equation}\label{c1}                       % c1 36
    \omega^C \;=\; x^{CC'}\, \pi_{C'} \;.
\end{equation}
Conjugation in Euclidean spacetime is
\begin{equation}\label{c2}                       % c2 37
    \hat{\pi}_1 \;=\; - \bar{\pi}_2 \;, \qquad
    \hat{\pi}_2 \;=\; + \bar{\pi}_1 \;.
\end{equation}
Twistors are defined only up to a scale. If we fix
this by $<\!\pi \hat{\pi}\!>\,=\,1\;$, we can identify
\begin{equation}\label{c3}                       % c3 38
    \pi_{C'} \;=\; u^+_{C'} \;, \qquad
    \hat{\pi}_{C'}  \;=\;  - u^-_{C'} \;,
\end{equation}
to get (\ref{a1})-(\ref{a2}) above. Then
$\;\omega^C = y^{+C}\;$ by (\ref{a7}). 
Next, eq.(5) of Ref.\cite{boels7} becomes 
by (\ref{a3}) and (\ref{a9}) above
\begin{equation}\label{c4}                      % c4 39
    \bar{\partial}_0 \;=\; D^{++}\;, \qquad
    \bar{\partial}_\alpha \;=\; \partial^+_C \;.
\end{equation}
Their action (eq.(16) of Ref.\cite{boels7}) is
\[                                              % c5 40
    S = \frac{1}{2} \int d^4x dk \,B_0
        (\bar{\partial}^\alpha A_\alpha +
                   g A^\alpha A_\alpha)
       + B^\alpha (\bar{\partial}_\beta A_0
       - \bar{\partial}_0 A_\beta + g [A_\beta, A_0])
\]
\begin{equation}\label{c5}
       - \frac{1}{4} \int d^4x dk_1 dk_2 \;H_1^{-1}
         B^0(\pi_1) H_1 H_2^{-1} B^0 (\pi_2) H_2
         \;<\pi_1 \pi_2>^4 .
\end{equation}
It can now be matched to (\ref{b12}) above.
Clearly $A_0 = A^{++},\; B_0 = B^{--}$ 
and the first two terms agree.
The third term has the same general structure, but
what is $H$? ``$H$ is a holomorphic frame of the gauge
bundle over $p^{-1}(x)$ such that the covariant
derivative of it vanishes on the sphere.''
I can't translate this hieroglyph (it looks
upside down), but if we identify
\begin{equation}\label{c6}                      % c6 41
     H_1\,H_2^{-1}\;=\;\mathcal{G}^{--}(u,v)\;,
\end{equation}
we get agreement since $<\!\pi_1\pi_2\!>\;=\;<uv> $.
Eq.(2.49) of Ref.\cite{boels8}, with notation
adjusted to match \cite{boels7}, gives their
expansion in MHV vertices 
\newpage
\[                                              %c7 42
    G^{(C'D')} G_{(C'D')} = - \sum_{n=2}^{\infty}
    (-i \sqrt{2} g)^{n-2} \int \sum_{p=2}^{n}
    tr( B_0^1 A_0^2 \cdots A_0^{p-1} B_0^p
    A_0^{p+1} \cdots A_0^p) 
\]
\begin{equation}\label{c7}
    \times \; <\pi_1 \pi_p>^4
    /(<\pi_1 \pi_2> <\pi_2 \pi_3> \cdots
      <\pi_n \pi_1>) ,
\end{equation}
which is exactly what results when (\ref{a19})
is substituted into the last term of (\ref{b12}).

Thus harmonics are much faster and clearer
than twistors. The trick is to start with
distinct gauge theories on $R_4$ and $S_2$ ,
and then mix them by a gauge transformation.  
The $CP_3$ apparatus just 
confused the issue. Now that a simple 
intuitive proof of Mason's Lagrangian exists, 
I hope to extend it to gravity using \emph{local}
twistors (heresy in Oxford). The powerful 
formalism of Ref.\cite{harmonics} could 
be generalized to any symmetric space,
and might have more applications.

\end{document}